\documentclass[conference]{IEEEtran}
\usepackage{cite}
\usepackage{amsmath,amssymb}
\usepackage{booktabs}
\usepackage{placeins}
\usepackage{dblfloatfix}
\usepackage{microtype}
\usepackage{tabularx}
\usepackage{array}
\usepackage{listings}
\usepackage{xcolor}
\usepackage{tikz}
\usepackage{url}
\usepackage[hidelinks]{hyperref}
\usetikzlibrary{positioning}
\newcolumntype{Y}{>{\raggedright\arraybackslash}X}

\lstdefinestyle{shell}{%
  language=bash,
  basicstyle=\ttfamily\footnotesize,
  backgroundcolor=\color{black!3},
  frame=single,
  rulecolor=\color{black!35},
  framerule=0.4pt,
  breaklines=true,
  breakatwhitespace=false,
  columns=fullflexible,
  keepspaces=true,
  showstringspaces=false,
  captionpos=t,
  abovecaptionskip=2pt,
  belowcaptionskip=5pt,
  aboveskip=8pt,
  belowskip=8pt
}
\title{Docker Containers vs. Virtual Machines:\\A Comparative Study of Architecture, Performance, Configuration, and Security}

\author{\IEEEauthorblockN{Faraz Gurramkonda, Akanksha Malla, Sayma Tamboli, and Shayesta Nazneen}
\IEEEauthorblockA{Department of Computer and Information Science\\
University of Michigan--Dearborn, Dearborn, MI, USA\\
\{gfaraz, amalla, saymat, shayesta\}@umich.edu}}

\begin{document}
\maketitle

\begin{abstract}
Modern application platforms must isolate workloads while preserving deployment speed, portability, resource efficiency, and security. Virtual machines (VMs) and Docker containers address this requirement at different abstraction layers: VMs virtualize hardware and run independent guest operating systems, whereas containers isolate processes while sharing the host kernel. This paper presents a comparative, literature-based analysis of the two approaches across architecture, configuration and lifecycle management, performance, scalability, and security. Published studies generally associate containers with shorter startup times, smaller images, higher workload density, and near-native execution for many workloads. These benefits depend on workload characteristics, storage and network drivers, resource controls, and experimental design. VMs introduce greater overhead but offer independent kernels, heterogeneous guest operating systems, and a stronger isolation boundary. The comparison therefore treats efficiency and isolation as a design trade-off rather than declaring one technology universally superior. A hybrid architecture, in which containers run inside hardened VMs, often provides a practical balance for cloud and multi-tenant systems.
\end{abstract}

\begin{IEEEkeywords}
containerization, Docker, virtual machines, virtualization, hypervisors, performance evaluation, cloud computing, operating-system security
\end{IEEEkeywords}

\section{Introduction}
Modern software systems frequently run multiple applications on shared infrastructure. Two dominant isolation mechanisms are hardware virtualization, commonly implemented through virtual machines, and operating-system-level virtualization, commonly implemented through containers. Both improve consolidation and reproducibility, but they differ in trust boundaries, resource overhead, startup behavior, portability, and operational workflow.

A VM exposes virtual hardware to a guest operating system. Each VM contains an independent kernel, system services, libraries, and application processes. Docker packages an application with its user-space dependencies as an image and executes the resulting container as an isolated set of processes on a shared host kernel. Linux namespaces separate process identifiers, networks, mounts, users, and other resources, while control groups account for and limit CPU, memory, and I/O consumption \cite{docker-security,nist-containers}.

This paper reorganizes the original course report into a research-paper structure. It provides a layered architectural comparison, a concise account of configuration workflows, a synthesis of reported performance evidence, a threat-oriented security analysis, and a practical technology-selection framework. The analysis is literature-based and does not claim a new controlled benchmark. Numerical results are treated as findings of the cited studies because virtualization performance varies with hardware, software versions, workload type, and configuration.

\section{Architectural Foundations}

\subsection{Virtual-Machine Architecture}
A VM stack normally contains physical hardware, a host or bare-metal hypervisor, virtual hardware, a guest operating system, guest libraries, and applications. A Type-1 hypervisor executes directly on hardware, whereas a Type-2 hypervisor runs above a host operating system. In both cases, the guest kernel owns the operating-system view inside the VM. This separation permits heterogeneous guest operating systems and provides a strong boundary between workloads, but it duplicates kernels and background services and therefore increases memory, storage, and boot overhead.

\subsection{Docker-Container Architecture}
Docker uses operating-system-level virtualization. The Docker Engine manages images, containers, networks, volumes, and the container lifecycle. An image contains application binaries and user-space dependencies in reusable layers. At runtime, containers share the host kernel but receive isolated views of selected kernel resources. This design avoids a separate guest kernel per application and enables fast creation, high density, and portable packaging. The trade-off is that every container depends on the security and compatibility of the shared kernel.

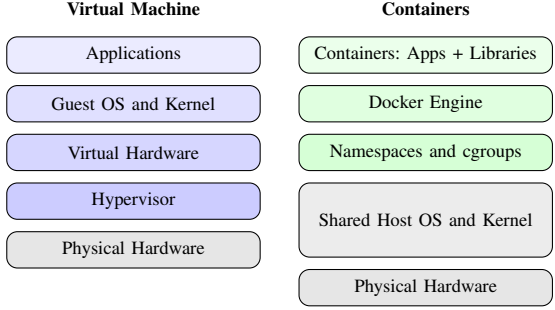
\begin{figure}[!t]
\centering
\begin{tikzpicture}[box/.style={draw,rounded corners,minimum width=3.35cm,minimum height=.48cm,align=center,font=\scriptsize},node distance=1.5mm]
\node[box,fill=blue!8] (a) {Applications};
\node[box,fill=blue!12,below=of a] (g) {Guest OS and Kernel};
\node[box,fill=blue!16,below=of g] (v) {Virtual Hardware};
\node[box,fill=blue!20,below=of v] (h) {Hypervisor};
\node[box,fill=gray!20,below=of h] (p) {Physical Hardware};
\node[font=\scriptsize\bfseries,above=1.5mm of a] {Virtual Machine};

\node[box,fill=green!8,right=5mm of a] (c) {Containers: Apps + Libraries};
\node[box,fill=green!12,below=of c] (r) {Docker Engine};
\node[box,fill=green!16,below=of r] (i) {Namespaces and cgroups};
\node[box,fill=gray!15,below=of i,minimum height=.98cm] (k) {Shared Host OS and Kernel};
\node[box,fill=gray!20,below=of k] (q) {Physical Hardware};
\node[font=\scriptsize\bfseries,above=1.5mm of c] {Containers};
\end{tikzpicture}
\caption{Conceptual comparison of VM and container layers.}
\label{fig:architecture}
\end{figure}

\begin{table*}[!tbp]
\caption{Architectural Comparison}
\label{tab:architecture}
\centering
\footnotesize
\setlength{\tabcolsep}{4pt}
\renewcommand{\arraystretch}{1.12}
\begin{tabularx}{\textwidth}{@{}p{2.25cm}YY@{}}
\toprule
\textbf{Dimension} & \textbf{Virtual Machine} & \textbf{Docker Container} \\
\midrule
Virtualization layer & Hardware abstraction managed by a hypervisor & Operating-system-level isolation managed by a runtime \\
Operating system & Independent guest kernel and user space & Shared host kernel; packaged application user space \\
Image composition & Guest OS, services, libraries, and applications & Application, libraries, configuration, and filesystem layers \\
Typical footprint & Often hundreds of MB to several GB & Often tens or hundreds of MB, depending on the image \\
Startup model & Guest operating-system boot & Isolated process creation from an image \\
OS flexibility & Supports heterogeneous guest operating systems & Requires compatibility with the host-kernel family \\
Isolation boundary & Separate guest kernels and virtual hardware & Process isolation within a shared kernel \\
Typical strength & Compatibility and strong separation & Portability, density, and rapid deployment \\
\bottomrule
\end{tabularx}
\end{table*}

\section{Configuration and Lifecycle Management}

\subsection{Container Workflow}
A container workflow describes the environment declaratively through an image definition, stores configuration separately from the image, and creates disposable runtime instances. Images can be versioned and distributed through registries, supporting consistent execution across development, test, and production environments. Listing~\ref{lst:docker-lifecycle} presents an abbreviated Docker Swarm configuration lifecycle based on the workflow discussed in the original report.

\begin{lstlisting}[style=shell,caption={Docker Swarm configuration lifecycle},label={lst:docker-lifecycle}]
printf '%s\n' 'This is a config' |
  docker config create my-config -

docker service create --name redis \
  --config my-config redis:alpine

docker service ps redis
docker config inspect my-config
docker service update --config-rm my-config redis
docker service rm redis
docker config rm my-config
\end{lstlisting}

Configuration objects should not contain secrets; secret-management facilities should be used for credentials and keys. Services and configuration objects should be removed after experiments to prevent stale resources.

\subsection{Virtual-Machine Workflow}
Configuring a VM generally requires selecting a hypervisor, creating virtual hardware, allocating CPU, memory, storage, and networking, installing a guest operating system, applying guest tools and updates, and configuring firewalls and access controls. Snapshots provide recovery points, while templates reduce repeated installation effort. Infrastructure-as-code can automate the process, but the deployed unit remains larger and more stateful than a container image.

Containers usually integrate naturally with continuous integration and continuous deployment because an immutable image can move through multiple environments. VMs remain valuable when applications require a different kernel, an older operating system, specialized virtual hardware, or a stronger workload boundary. Cloud platforms commonly combine the models by running orchestrated containers inside VM-based worker nodes.

\section{Performance Evidence}

\subsection{Method}
The original report considered CPU computation, memory throughput, storage I/O, web-request throughput, image size, startup time, and operation speed, with tools including Sysbench, Phoronix Test Suite, ApacheBench, IOzone, Intel Linpack, and Glances. This revision uses those dimensions as a framework for synthesizing published evidence rather than presenting a new experiment. Results from different publications are not pooled because their hardware, hypervisors, operating systems, Docker versions, workloads, and resource allocations are not identical.

\subsection{Synthesis of Reported Results}
Prior studies commonly observe lower overhead for containers, particularly for startup, image footprint, and memory use \cite{yadav2018,raj2020,potdar2020,felter2015}. Raj and Nath reported VM start and stop times measured in seconds, compared with millisecond-scale container operations for a prebuilt ReactJS environment \cite{raj2020}. Their Docker image was also substantially smaller than the corresponding VM image. Potdar \emph{et al.} evaluated CPU, memory, storage, load, and execution-time behavior using standard benchmark suites and reported favorable Docker results across several tested workloads \cite{potdar2020}. Yadav \emph{et al.} similarly reported lower request-execution times for Docker than for a KVM-based VM in their environment \cite{yadav2018}.

These findings should not be generalized as a fixed percentage advantage. Containers may approach native performance because processes use the host kernel directly, while VMs introduce guest-kernel and virtual-device layers. Performance nevertheless changes with hardware-assisted virtualization, paravirtualized drivers, storage back ends, networking modes, NUMA placement, CPU pinning, memory overcommit, and orchestration overhead. Big-data experiments also show that application tuning and resource configuration can be as important as the virtualization mechanism \cite{zhang2018}.

\begin{table*}[!tbp]
\caption{Performance Interpretation by Dimension}
\label{tab:performance}
\centering
\footnotesize
\setlength{\tabcolsep}{4pt}
\renewcommand{\arraystretch}{1.12}
\begin{tabularx}{\textwidth}{@{}p{2.15cm}YYp{3.1cm}@{}}
\toprule
\textbf{Dimension} & \textbf{Expected Container Behavior} & \textbf{Expected VM Behavior} & \textbf{Important Controls} \\
\midrule
Startup and shutdown & Process creation from an existing image; typically fast & Full or partial guest-OS boot and shutdown & Warm/cold state, image cache, initialization services \\
CPU & Often near native for compute-bound workloads & Small to workload-dependent virtualization overhead & CPU pinning, frequency scaling, vCPU allocation \\
Memory & No duplicated guest kernel per container & Guest kernel and services consume additional memory & Page cache, ballooning, limits, overcommit \\
Storage I/O & Can be fast, but layered filesystems may add overhead & Depends on virtual disk and paravirtualized driver & Cache mode, filesystem, volume driver, record size \\
Network & Bridge, NAT, and overlay choices affect latency & Virtual NIC and switch affect latency & MTU, offload, bridge or host mode, driver \\
Density & More instances per host are usually practical & Lower density because each VM contains an OS & Service footprint, reservation, isolation target \\
\bottomrule
\end{tabularx}
\end{table*}

A credible future benchmark should use identical hardware, pin equal CPU and memory resources, record software versions, repeat trials, report confidence intervals, distinguish cold from warm starts, and publish scripts and raw measurements.

\section{Security Analysis}

\subsection{Container Security}
Namespaces and cgroups provide useful isolation and resource governance, but they do not create an independent kernel. Container security depends on the host kernel, runtime, daemon configuration, image provenance, and application privileges. Important threats include vulnerable or malicious images, an exposed Docker daemon, excessive Linux capabilities, privileged containers, unsafe bind mounts, unrestricted inter-container traffic, weak secret handling, and unpatched kernel vulnerabilities \cite{docker-security,nist-containers}.

Mitigations include minimal and trusted base images, image signing and scanning, non-root execution, rootless mode where feasible, removal of unnecessary capabilities, seccomp and mandatory-access-control profiles, read-only filesystems, network segmentation, resource limits, protection of the daemon socket, and rapid patching of the host and runtime.

\subsection{Virtual-Machine Security}
VMs place an independent guest kernel between the application and hypervisor, generally strengthening tenant isolation. They remain exposed to hypervisor vulnerabilities, VM escape, insecure virtual-device implementations, side-channel attacks, vulnerable images, stale snapshots, weak management interfaces, and VM sprawl. Mitigations include minimizing and patching the hypervisor, isolating management networks, hardening guest images, encrypting sensitive storage, restricting administrative APIs, monitoring guest and hypervisor logs, and enforcing lifecycle policies.

\begin{table*}[!tbp]
\caption{Security Trade-offs and Controls}
\label{tab:security}
\centering
\footnotesize
\setlength{\tabcolsep}{4pt}
\renewcommand{\arraystretch}{1.12}
\begin{tabularx}{\textwidth}{@{}p{2.15cm}YY@{}}
\toprule
\textbf{Area} & \textbf{Docker Containers} & \textbf{Virtual Machines} \\
\midrule
Primary boundary & Namespaces, capabilities, cgroups, and kernel security modules & Hypervisor, virtual hardware, and independent guest kernels \\
High-impact compromise & Host-kernel or daemon compromise can affect multiple containers & Hypervisor compromise or VM escape can affect multiple guests \\
Artifact risk & Vulnerable images, embedded secrets, untrusted registries & Vulnerable templates, stale snapshots, unpatched guest images \\
Network risk & Permissive bridges, published ports, overlay misconfiguration & Virtual-switch, management-plane, and segmentation errors \\
Preferred controls & Least privilege, image scanning/signing, seccomp, MAC profiles & Hypervisor hardening, guest patching, management isolation, inventory \\
Best fit & Trusted or moderately trusted workloads with strong hardening & Untrusted tenants, distinct kernels, high-assurance separation \\
\bottomrule
\end{tabularx}
\end{table*}

No virtualization technology is secure by default. The appropriate comparison is between hardened configurations under a defined threat model. NIST recommends addressing risks across images, registries, orchestrators, containers, and host operating systems rather than relying on isolation alone \cite{nist-containers}.

\section{Technology-Selection Framework}
Selection should begin with workload and risk requirements rather than a universal preference.
\begin{itemize}
  \item \textbf{Prefer containers} when rapid scaling, high density, immutable delivery, microservices, and CI/CD speed dominate, and workloads can share a compatible host kernel.
  \item \textbf{Prefer VMs} when workloads require different operating systems or kernels, include legacy dependencies, demand stronger tenant boundaries, or require specialized virtual hardware.
  \item \textbf{Prefer a hybrid design} when teams need container portability and orchestration together with VM-level tenant, node, or regulatory boundaries.
\end{itemize}

Cost comparisons should include not only instance density but also operational tooling, staff expertise, patching, monitoring, backup, licensing, incident response, and compliance. A container's smaller footprint does not automatically make it the correct unit for every stateful or security-sensitive workload.

\section{Discussion}
The evidence supports three broad observations. First, containerization reduces duplicated operating-system components and therefore improves startup speed and workload density in many environments. Second, VM overhead has declined with hardware-assisted virtualization and optimized drivers, making the difference workload-dependent rather than absolute. Third, isolation and performance are related: the shared kernel that gives containers their efficiency also concentrates risk at the host boundary.

Several limitations constrain this comparison. The reviewed experiments use different hardware and software versions, and some measurements compare a warm, prebuilt container with a cold VM boot. Benchmark tools approximate only selected production behaviors. Long-running services may care more about tail latency, noisy-neighbor interference, failure recovery, and operational control than initial startup time. Future work should evaluate containers, VMs, and containers-in-VMs on identical modern platforms using reproducible workloads and security configurations.

\section{Conclusion}
Docker containers and virtual machines solve related isolation problems at different layers. Containers package applications efficiently, start quickly, and support high-density, portable deployment. VMs consume more resources but provide independent guest kernels, heterogeneous operating-system support, and a stronger default isolation boundary. Published performance studies generally favor Docker for startup, image size, memory footprint, and many application workloads, but the magnitude of improvement is configuration- and workload-specific. Security likewise depends on implementation and hardening rather than the technology label alone. For many production environments, the most practical architecture is not an exclusive choice: containers provide the application-delivery unit, while VMs define infrastructure and trust boundaries.

\FloatBarrier

\end{document}